# ON THE PRODUCT SELECTIVITY IN THE ELECTROCHEMICAL REDUCTIVE CLEAVAGE OF LIGNIN MODEL COMPOUNDS

Marcia Gabriely A. da Cruz[a, *], Bruno V. M. Rodrigues[a], Andjelka Ristic[b], Serhiy Budnyk[b], Shoubhik Das[c], Adam Slabon[a,*]

[a]Department of Materials and Environmental Chemistry, Stockholm University, Stockholm, Sweden; [b]AC²T research GmbH, Wiener Neustadt, Austria; [c]ORSY Division, Department, of Chemistry, Universiteit Antwerpen, Antwerpen, Belgium.

**ABSTRACT:** Research towards the production of renewable chemicals for fuel and energy industries has found lignin valorization as key. With a high carbon content and aromaticity, a fine-tuning of the depolymerization process is required to convert lignin into valuable chemicals. In context, model compounds have been used to understand the electrocatalyzed depolymerization for mimicking the typical linkages of lignin. In this investigation, 2-phenoxyacetophenone, a model compound for lignin β-O-4 linkage, was electro-catalytically hydrogenated (ECH) in distinct three-electrode setups: an open and a membrane cell. A deep eutectic solvent based on ethylene-glycol and choline chloride was used to pursue sustainable routes to dissolve lignin. Copper was used as electrocatalyst due to the economic feasibility and low activity towards hydrogen evolution reaction (HER), a side reaction of ECH. By varying the cell type, we demonstrate a simple ECH route for the generation of different monomers and oligomers from lignin. Gas chromatography of the products revealed a higher content of carbonyl groups in those using the membrane cell, whereas the open cell produced mostly hydroxyl-end chemicals. Aiming at high value-added products, our results disclose the cell type influence on electrochemical reductive depolymerization of lignin. This approach encompasses cheap transition metal electrodes and sustainable solvents.

## INTRODUCTION

Lignin depolymerization as a strategy to add value to the most available natural carbon source worldwide has been a challenge since the 1930s, when the earlier findings in the field were published[1,2]. Initially, this macromolecule was treated as a feedstock like oil and coil, making use of similar conversion techniques[3]. Deeper investigations on the structure revealed its intricate interconnectedness, which leads to a complex three-dimensional structure containing diverse functional groups. Its recalcitrance can be explained by the presence of alkyl and aryl units connected by strong C-O and C-C bonds forming specific types of linkage, such as β-O-4, α-O-4, β-β, β-5, 5-5, and 4-O-5. Among them, the β-O-4 bond is the most common, which is responsible for approximately 50% of the macromolecule linkage.

Ideally, an efficient depolymerization process should produce functionalized aromatic compounds in sufficient yield at room temperature and selectivity in order to enable extraction and further valorization[4,5]. Among different catalytic methods to depolymerize lignin, electrocatalysis has become more widely recognized as an environment-friendly process with several benefits[6–11]. This method i) allows to use clean energy as electricity source, ii) enhances catalyst activity and selectivity by controlling the cell potential, which also avoids the use of harsh reaction conditions (high temperature, pressure, and harmful solvents), iii) makes the product-catalyst separation step unnecessary, as electrodes act as heterogeneous catalysts[4, 10–12].

In general, electrocatalytic oxidation (ECO) is the most reported approach due to the possibility of carbonylation of α-hydroxyl groups, which are commonly present in the lignin chain[6, 13–17]. The presence of α-carbonyl can promote bond cleavage and inhibit condensation[4, 11]. However, in this type of process, the occurrence of overoxidation towards organic acids and $CO_2$ is a big challenge[18]. In addition, electrocatalytic hydrogenation (ECH) manage to be an alternative to thermal catalytic hydrotreating process. The reduction decreases oxygen content in the end products bringing advantages to lignin oil's generation process with regard to sustainability and energetic costs[19–21]. The interesting phenolic products obtained by catalytic hydrogenation/hydrogenolysis via heterogeneous or enzymatic catalysis[22–24] makes the ECH a way to overthrow the majority of the problems found in the previous methods. This is due to the possibility to increase the efficiency and sustainability of a process that can yield similar products.

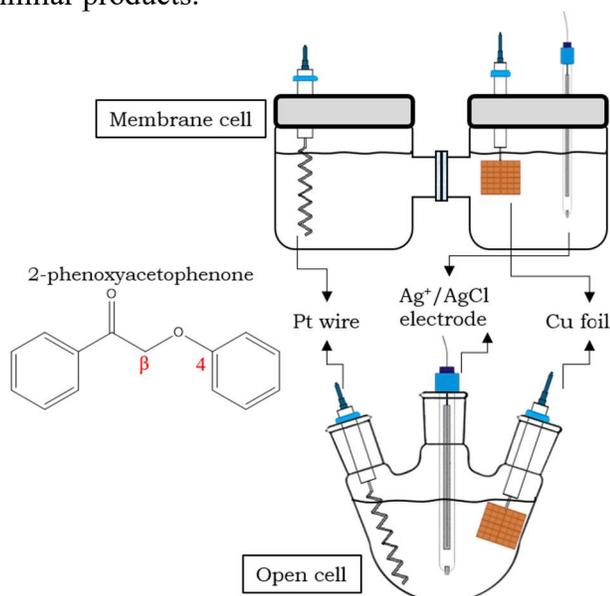

Figure 1. (a) Chemical structure of 2-phenoxyacetophenone (2-PAP) with the β-O-4 linkage highlighted and (b) scheme of the membrane and open electrolytic system.

First attempts of novel reactions and systems using any type of technical lignin can be challenging due to the inherent heterogeneity of this macromolecule. Model compounds that mimic its typical linkages are widely used to enlighten and to provide a better understanding of the electrocatalyzed depolymerization of lignin[2, 10, 25–29]. In this investigation, we draw upon the same procedure to evaluate simple, scalable, sustainable, and economically feasible setups to promote ECH of a lignin model compound. The cleavage of the β-O-4 bond of 2-phenoxyacetophenone (2-PAP) was chosen as a simple model compound for the lignin (Figure 1). Therefore, we disclose a better elucidation of the mechanism and selectivity of lignin depolymerization in distinct electrochemical setups.

ECH has the drawback of simultaneously promoting the hydrogen evolution reaction (HER) as a competing reaction during organic hydrogenation on a metallic catalyst. According to the hydrogen binding energy (HBE) theory, the best-known electrocatalysts for the HER reaction are noble metals (Pt, Ru, and Ir)[30–32]. In order to decrease the HER activity and balance it with lignin depolymerization, we evaluated the materials that presented lower activity towards HER, considering also value and availability. Therefore, Cu was chosen as our working electrode[31, 33, 34].

Due to the previously documented ability to dissolve lignin, deep eutectic solvents (DES)[35–37] represent a promising choice as sustainable solvents[38, 39]. Herein, we used a mixture composed of ethylene glycol and choline chloride with a molar ratio of 2:1 as a deep eutectic solvent for lignin. Water was added in $10\%_{vol.}$ to increase the electrolytic activity and to provide hydrogen; the latter as a result of the accompanying HER. Water also improves the solution viscosity, which consequently decreases possible mass flow limitations.

We proceeded with the depolymerization reactions in two different three-electrode setups: an open cell and a membrane cell (Figure 1). The comparison of both reactors could help to determine how the presence of an anodic reaction occurring on the counter electrode can influence the ECH. We used the same reaction parameters for both systems.

## EXPERIMENTAL

### General materials

2-phenoxyacetophenone (98%) and choline chloride (98%), were purchased from Tokyo Chemical Industry (TCI). Ethylene glycol

(analytical grade), and ethyl acetate (technical grade) were purchased from Sigma-Aldrich. All materials were used as received.

The copper foil was purchased from WAT Venture, Poland, and it was ultrasonically washed with $H_2SO_4$ solution, followed by DI water rinsing prior to the reaction.

### ECH system and experimental apparatus

DES was prepared by mixing choline chloride and ethylene glycol (ChCl-EtGl) in a 1: 2 molar ratio by magnetic stirring over a hot plate. Afterward, 3 g.L$^{-1}$ of 2-PAP was dissolved in the DES by mechanical stirring at 500 rpm and an aqueous fraction of 10%$_{vol.}$ was then added.

The electrochemical depolymerization was performed in an undivided cell and a membrane cell using a potential of –1.7 V in the presence of 1 cm$^2$ of Cu foil electrode area for 20 h. A platinum wire and Ag/AgCl electrode were the counter and reference electrodes, respectively.

### Product recovery and extraction

The extraction process of the depolymerization products started with the precipitation of insoluble fractions followed by liquid-liquid extraction (LLE) of soluble fractions. The first step was conducted by diluting the reaction solution with 1 M $H_2SO_4$ until pH 2, which has not led to precipitates. Ethyl acetate was used as an organic extractant phase and added to the aqueous sulphuric acid solution to proceed with the LEE process. The mixture was stirred at 250 rpm overnight in an orbital shaker. Two phases were obtained and separated using a separating funnel. The liquid fraction of products was then extracted from the organic phase in a rotary evaporator.

### Characterization of products

GC-MS samples were dissolved in a methanol–chloroform mixture (volumetric ratio 3:7, dilution factor of 1:1000). Sample solutions were then introduced into a LTQ Orbitrap XL hybrid tandem high-resolution mass spectrometer from Thermo Fisher Scientific (Bremen, Germany) by direct infusion applying a flow rate of 5 µL/min. The instrument was fitted with electrospray ionization (ESI) ion source and operated in positive or negative ion mode respectively. Nitrogen was used as sheath gas. Helium was used both as buffer and collision gas in the linear ion trap section where lower energy collision-induced dissociation (CID) was performed. For the identification of chemical structures by tandem MS, product ions were generated in the linear ion trap via CID and detected by the high-resolution orbitrap section of the instrument at a resolution of 60,000 (full width at half maximum, FWHM). The mass measurements were acquired with a mass accuracy of 5 ppm or better. Data processing and interpretation were done using the software tools Xcalibur version 2.0.7 and Mass Frontier version 6.0 from Thermo Fisher Scientific (Bremen, Germany). m/z describes the mass-to-charge ratio of the detected ions. As all negatively charged ions analyzed were single charged species, m/z also referred to the monoisotopic molecular masses of the detected ions. It is noted that ionization capability of chemical compounds is dependent on the chemical structure. Thus, for spectrum interpretation, especially of the different depolymerization routes and model products, the intensities detected can only be compared between the same species found in the analyzed samples obtained by the similar procedure but not between different species.

FT-IR analysis was performed using a Varian 610-IR FT-IR spectrometer. The spectra were collected in the range of 400-4000 cm$^{-1}$ with 16 scans at 4 cm$^{-1}$ resolution and 1 cm$^{-1}$ interval at room temperature.

$^1$H NMR spectra were recorded on a Bruker DRX 4OO NMR instrument at 25 °C in DMSO-$d_6$ containing tetramethylsilane as an internal standard. An inversed gated proton decoupling with 90⁰ pulse angle, and delay time of 10 s was applied. The spectra were recorded in 32 scans with a 10 s delay.

## RESULTS AND DISCUSSION

Chronoamperometry gave similar constant current densities for both reactors, $I_{av}$ of –14 and –16 mA.cm$^{-2}$ for open and membrane cells, respectively (Figure 2). However, the color difference between the solutions clearly reveals the difference between both setups. Studies on lignin depolymerization relate similar color change toward a clear solution when using anodic current[40].

For the 2-PAP solutions in DES, a similar behavior was observed after 10 h, becoming yellow and translucent later in the open cell and darker in the membrane cell (Figure 3). At first, the translucent solution suggested a rearrangement toward the initial structure, but further characterization could explain the color by the higher concentration of alcohols in the open cell instead of ketones and ethers in the membrane cell, which confers the darker color for the solution.

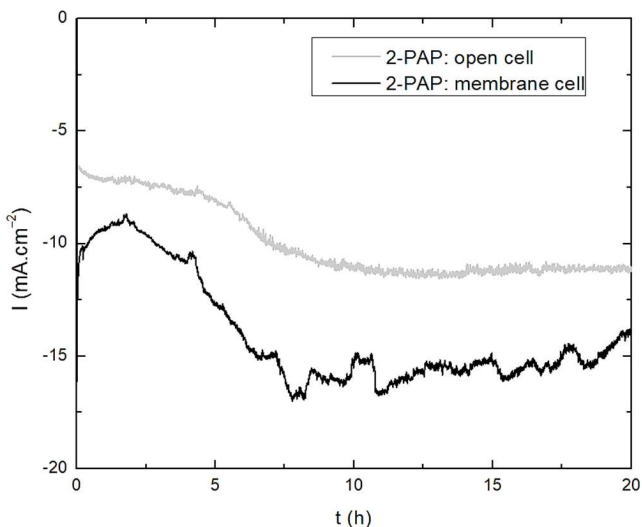

Figure 2. Constant potential ECH of 2-PAP at – 1,7V in deep eutectic solvent for open and membrane electrochemical cell.

The identification of products arising from the depolymerization of lignin is a complex process. From a simple model compound, one could expect an elementary blend of products. However, the GC-MS revealed a wide spectrum, which can be attributed to the low sensitivity of the electrochemical process, making the quantification even more difficult (Figure 3). Monomers and oligomers were observed as products, the latter from a likely coupling arrangement due to the long-term reaction.

Despite the diversity of the GC-MS spectrum, a noticeable pattern that differs in each type of reactor was noted. The products obtained from the membrane cell have a higher content of carbonyl groups, while those obtained from the open cell produced a higher concentration of hydroxyl ends. The oxygen evolution reaction (OER) occurring in the same chamber may favor ether cleavage, which increases the content of alcoholic ends. A similar pattern was not observed on the membrane cell, where the ether bonds and some carbonyl groups were preserved, and the alcoholic ends came from the carbonyl reduction of the ketone present in the model compound.

Lignin structural characterization was managed by the combination of NMR and IR spectroscopy techniques, determining the functional groups present in the macromolecule[41]. Associating the techniques with the presented GC-MS result confirmed the pattern in the distribution of the products. The NMR spectra (Figure 4) revealed the presence of phenolic groups in both systems, the peaks highlighted at ~2 ppm and ~4 ppm are stronger for the membrane cell, evidencing the ether and ketone terminations. The IR spectra (Figure 5) reinforce their presence showing them bonded to aryl groups, as suggested by the GC-MS. It can be seen from the combination of two peaks at 1719-1240 $cm^{-1}$ for ketones and 1377-1038 $cm^{-1}$ for ether linked to aryl terminations.

The formation of acetophenone and pinacol was observed in a previous study when a current of 5 mA was applied to a 2-PAP solution in dimethylformamide for 2.5 h, in a membrane cell[27]. Although the cited electrocatalytic cleavage had different reaction parameters, it shows that when only current is applied without a mediator, the model compound can keep the ketone termination and also suffer reductive dimerization, like some of the compounds here presented. Analogous comportment was observed for oxidized lignin models in a divided cell[2]. In an open cell, the electroreductive depolymerization of model compounds promoted by sodium borohydride also led to phenolic terminations after cleavage of 4-O-5, α-O-4, and β-O-4 model compounds of lignin[26].

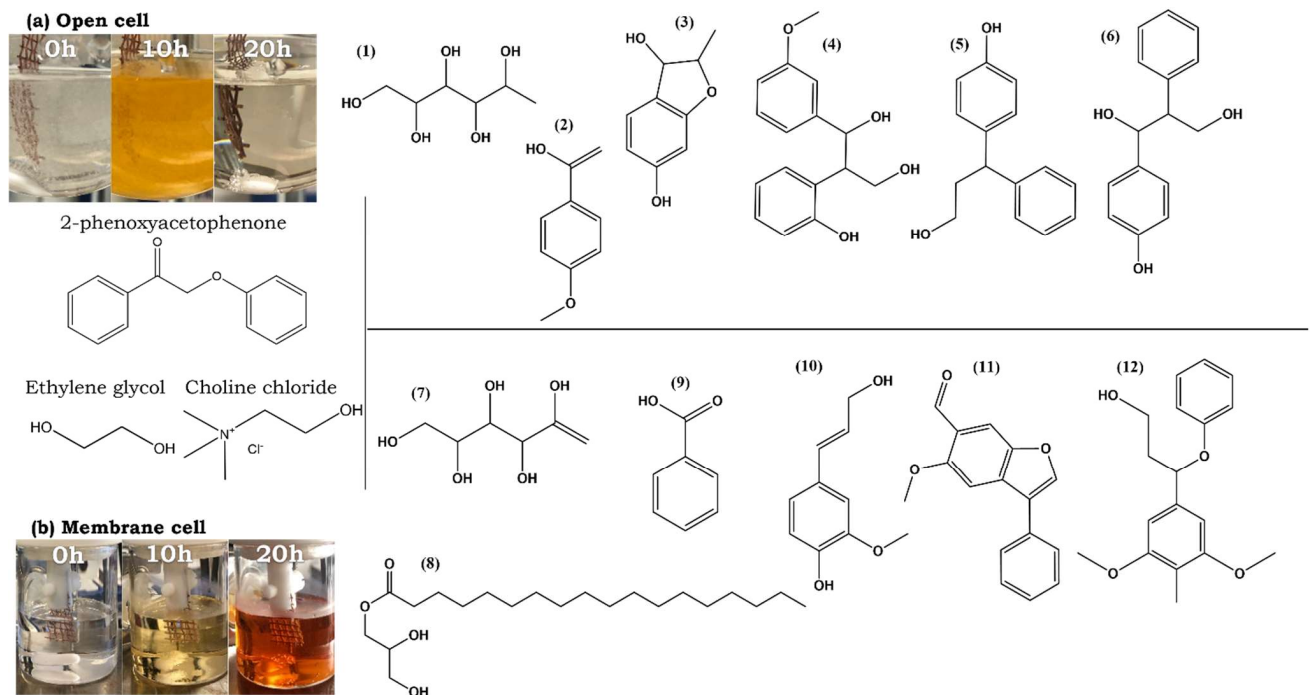

Figure 3. Color change of the 2-PAP solution's color for (a) open and (b) membrane cell and the main identified products using GC-MS (1-12).

DES has emerged as a promising new class of green solvent for lignin processing, but a proper understanding of the interaction between the solvent and the macromolecule has not been achieved yet[36]. Herein, for both cell types, we could observe that the model compound interacted with the DES, primarily with the ethylene glycol. Based on GC-MS analysis, some of the products showed an interaction of the aromatic rings and the hydrocarbon chain of ethylene glycol present in the solvent. Thus, even with consecutive extractions, it was not possible to isolate the products from the DES. Different from ionic liquids that count on ionic interaction during their synthesis process, the DES is a mixture of a hydrogen bond acceptor and donor and the strong intermolecular force is responsible for its stability[42]. In the presence of considerable cathodic potential, the intermolecular interaction can be overdue by the reaction with 2-PAP.

At first, we aimed to evaluate the possibility to use Cu as working electrode with a Cu foil. The Cu electrode showed the desired activity for the reaction of the model compound even though the selectivity for β-O-4 cleavage was not evidenced. As previously discussed, other metals such as Ag or Au which are known to exhibit also low activity for the HER, could potentially substitute Cu for better selectivity towards β-O-4 cleavage [30, 43, 44]. Moreover, the activity of an electrode is strongly influenced by its geometry and area. Catalyst geometries like monodisperse nanocrystals could also modify the activity and selectivity of the metal electrode[45–48].

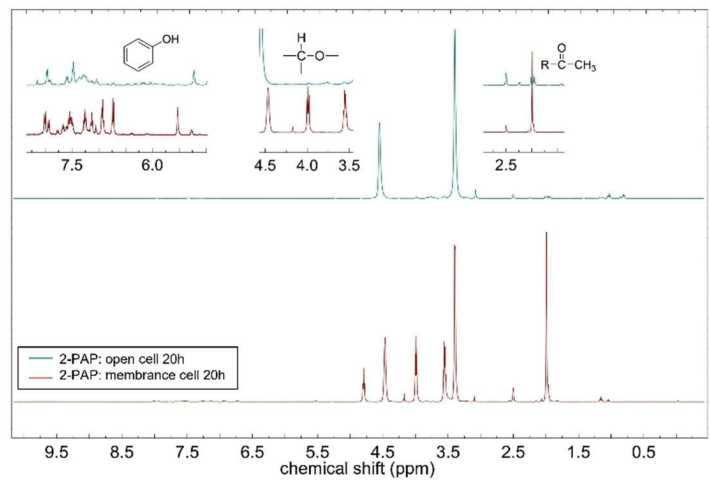

Figure 4. NMR spectra of ECH of 2-PAP products in open and membrane cells.

## CONCLUSIONS AND OUTLOOK

Electroreductive cleavage of β-O-4 linkage of 2-PAP was investigated as a model compound of lignin aiming to evaluate the effect towards the selectivity of obtained products in two electrochemical setups. In conclusion, we disclose a simple one-step and cheap electrocatalytic hydrogenation that can generate different types of products varying with the type of cell. Using Cu foil as electrocatalyst with its modest activity towards the HER enables electrochemical reductive cleavage of the model compounds, being as such equivalent to lignin depolymerization. Although the aimed selectivity for β-O-4 cleavage was not observed, we believe that this method is general for other transition metals with low activity for HER, like Ag and Au.

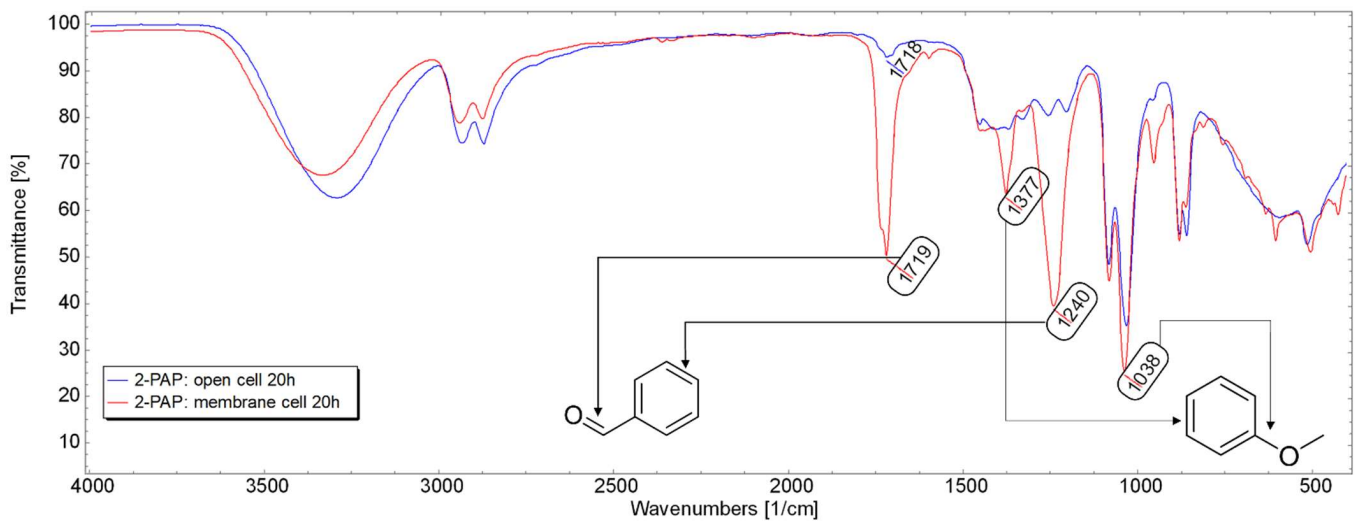

Figure 5. FTIR spectra of ECH of 2-PAP products in open and membrane cells.

The DES solvent presented partial interaction with the model compound. Even though a proper understanding of the behavior of this class of solvent in lignin dissolution is still lacking, herein our results indicate that the applied potential promotes an interaction between the DES and 2-PAP. Some of the obtained products were generated from the interaction of the ethylene glycol present on the DES with the model compound.

As 2-PAP is one of the simple representants of the β-O-4 linkage of lignin, further studies will be needed with more complex model compounds to determine the influence of the side groups affects the reactivity of the typical linkages. The critical advancement of this work is the demonstration of ECH on the model compound and the elucidation of the influence of reactor type on product selectivity. The results show that the membrane, or its absence, plays an important role concerning obtained depolymerization products. Further work on depolymerization of Kraft lignin is ongoing in our laboratory.

## ASSOCIATED CONTENT
### *AUTHOR INFORMATION*

*Corresponding Author*

Adam Slabon – Department of Materials and Environmental Chemistry, Stockholm University, 106 91 Stockholm, Sweden; orcid.org/0000-0002-4452-1831; email: adam.slabon@mmk.su.se

Márcia G. A. Cruz – Department of Materials and Environmental Chemistry, Stockholm University, 106 91 Stockholm, Sweden; orcid.org/0000-0002-0357-0012; email marciagabriely.alvesdacruz@mmk.su.se

*Authors*

Bruno V. M. Rodrigues – Department of Materials and Environmental Chemistry, Stockholm University, 106 91 Stockholm, Sweden; email bruno.manzolli@mmk.su.se

Andjelka Ristic – AC2T research GmbH, Viktor-Kaplan-Straße 2C, 2700 Wiener Neustadt, Austria; email: andjelka.ristic@ac2t.at

Serhiy Budnyk – AC2T research GmbH, Viktor-Kaplan-Straße 2C, 2700 Wiener Neustadt, Austria; email: serhiy.budnyk@ac2t.at

Shoubhik Das – Department of Chemistry, University of Antwerp, Groenenborgerlaan 171, 2020 Antwerpen, Belgium; orcid.org/0000-0002-4577-438X, email: shoubhik.das@uantwerpen.be

*Notes*

The authors declare that they have no known competing financial interests or personal relationships that could have appeared to influence the work reported in this paper.

**ACKNOWLEDGMENT**

This work was financially supported by the Carl Tryggers Stiftelse för Vetenskaplig Forskning under grant CTS 39: 194. We acknowledge additional financial support by the COMET InTribology project (FFG No. 872176, project coordinator: AC2T research GmbH). The work was carried out in collaboration with "Excellence Centre of Tribology" AC2T research GmbH.